\documentclass[twocolumn,tighten]{aastex701}
\usepackage{multirow} 
\usepackage{url}
\usepackage{booktabs}
\shorttitle{A Close-Pair Strong Lens at $z_d=0.79$}
\shortauthors{He et al.}
\submitjournal{ApJL}
\graphicspath{{./}{figures/}}
\begin{document}

\title{Discovery of a Strong-lens Galaxy Pair with the Smallest Projected Separation}

\author[0000-0001-8554-9163]{Zizhao He}
\affiliation{Department of Physics, Nanchang University, Nanchang, 330031, China}
\affiliation{Center for Relativistic Astrophysics and High Energy Physics, Nanchang University, Nanchang, 330031, China}
\affiliation{Purple Mountain Observatory, Chinese Academy of Sciences, Nanjing, Jiangsu, 210023, China}
\email{zzhe@ncu.edu.cn}

\author[0009-0009-9255-920X]{Limeng Deng}
\affiliation{Purple Mountain Observatory, Chinese Academy of Sciences, Nanjing, Jiangsu, 210023, China}
\affiliation{School of Astronomy and Space Sciences, University of Science and Technology of China, Hefei 230026}
\email{892667076@qq.com}

\author[0009-0006-9345-9639]{Qihang Chen}
\affiliation{School of Physics and Astronomy, Beijing Normal University, Beijing, 100875, China}
\affiliation{Institute for Frontier in Astronomy and Astrophysics, Beijing Normal University, Beijing, 102206, China}
\email{15950525352@163.com}

\author[0000-0002-9063-698X]{Yiping Shu}
\affiliation{Purple Mountain Observatory, Chinese Academy of Sciences, Nanjing, Jiangsu, 210023, China}
\email{yiping.shu@pmo.ac.cn}

\author[0000-0001-6800-7389]{Nan Li}
\affiliation{National Astronomical Observatories, CAS, Beijing 100101, China}
\email{nan.li@nao.cas.cn}

\author[0009-0003-6635-9593]{Di Wu}
\affiliation{National Astronomical Observatories, CAS, Beijing 100101, China}
\affiliation{School of Astronomy and Space Science, University of Chinese Academy of Sciences, Beijing 100049, China}
\email{wudi@bao.ac.cn}

\begin{abstract}
We present the spectroscopic confirmation and lens modeling of HSC~J0233$-$0205, a strong-lensing system produced by a close pair of elliptical galaxies at $z_d=0.790\pm0.022$ that lenses a multi-component background source at $z_s=2.160\pm0.002$. The two deflectors are separated by only $0\arcsec.481\pm0\arcsec.004$ ($3.596\pm0.046$ kpc), making this system a compact galaxy-pair lens at relatively high redshift. Joint five-band HSC lens modeling requires two mass components, with Einstein radii of $0\arcsec.774\pm0\arcsec.011$ and $0\arcsec.767\pm0\arcsec.014$, and yields a circularized Einstein radius of $\theta_{\rm E}=1\arcsec.549\pm0\arcsec.018$ for the overall system. The lensed source is reconstructed with three components: two extended components separated by $0\arcsec.462\pm0\arcsec.013$ ($3.830\pm0.108$ kpc), together with a compact component nearly aligned with one of them. Integrating the lensing convergence map within the critical curve gives a projected mass of $M_{\rm lensing,crit}=(9.626\pm0.010)\times10^{11}M_\odot$. Combining this with the stellar mass inferred from SED fitting, $M_{\ast,{\rm crit}}=(1.570\pm0.240)\times10^{11}M_\odot$, we obtain a projected dark-matter fraction within the critical curve of $f_{\rm dm}=83.7\pm2.5\%$. Within the $z$-band effective radii of the two deflectors, the corresponding dark-matter fractions are $82.1\pm4.5\%$ and $65.7\pm6.7\%$, respectively. HSC~J0233$-$0205 is therefore a compact, high-redshift galaxy-pair lens. Future high-resolution imaging and spatially resolved spectroscopy will enable detailed tests of merger signatures, the redistribution of stellar and dark matter, and possible light--mass offsets in the lens plane.
\end{abstract}


% Additionally, the lensing galaxy at a higher redshift is itself deflected by the other lensing galaxy at a lower redshift, forming an arc-shaped galaxy. ...... The highly magnified, arc-shaped background galaxy further enables detailed studies of its internal structure and star-formation activity. AGNs, which typically serve as sources, can also act as deflectors in some rare systems \citep[e.g.,][]{Millon2023}, providing a powerful probe of the co-evolution between the central black hole and its host through joint constraints from lensing and AGN physics. For instance, \cite{Dux2024} describes a `zig-zag' lens where the light from a QSO is deflected by two lenses simultaneously, resulting in a six-image configuration.  Such `zig-zag' systems provide a unique laboratory for multi-plane lensing, allowing simultaneous constraints on the mass distributions and dark-matter fractions of two deflectors and their halo interaction.

\section{Introduction} \label{sec:intro}

Strong gravitational lensing provides a direct measurement of projected mass on kiloparsec scales and, when combined with stellar-population modeling and/or stellar dynamics, enables the stellar and dark-matter contributions in galaxy centers to be separated \citep{Koopmans2006,Gavazzi2007,Bolton2008,Treu2010,Auger2010,Granata2023}. Galaxy-pair lenses are particularly powerful laboratories, because the mutual gravitational interaction between the two deflectors can modify their stellar and dark-matter distributions through tidal stripping, baryonic inflows, halo overlap, and the redistribution of material during the early stages of merging \citep{Toomre1972,Barnes1992,Mihos1996,Grillo2011,Lee2018}. These effects are expected to be most significant when the two galaxies are separated by only a few kiloparsecs, where their stellar envelopes and inner dark-matter halos may already strongly overlap \citep{Gonzalez2023}. In such systems, strong lensing can therefore address several key questions: how the total mass is shared between the two closely projected galaxies, whether their central dark-matter fractions differ from those of isolated early-type galaxies, and, with sufficiently high-resolution data, whether the peaks or centroids of the mass distribution are offset from the observed stellar light \citep{Keeton2003,Grillo2011,Shu2016b}.

The last point is particularly important for dark-matter physics. In a collisionless cold-dark-matter scenario, the dark matter, stars, and galaxies are generally expected to remain nearly coincident during an interaction, apart from perturbations induced by tides, projection effects, or modeling systematics \citep{Clowe2006,Randall2008,Harvey2015}. In contrast, if dark matter has a non-negligible self-interaction cross section, dark-matter halos may experience an effective drag during collisions or close encounters, potentially producing measurable offsets between the dark-matter distribution and the collisionless stellar component \citep{Spergel2000,Markevitch2004,Kahlhoefer2015,Kim2017,Robertson2017}. Searches for such light--mass offsets have therefore been widely used as tests of self-interacting dark matter, especially in cluster-scale mergers \citep{Markevitch2004,Clowe2006,Randall2008,Harvey2015,Massey2015,Massey2018}. Close galaxy-pair lenses provide a route to extending this type of test to much smaller physical scales, although robust offset measurements require imaging resolution and lens-model constraints beyond those available for many current systems \citep{Massey2015,Gilman2021}.

However, only a small number of strong lenses with close-pair deflectors have been reported to date, and systems with projected separations of only a few kiloparsecs remain especially rare \citep[e.g.,][]{Keeton2003,Grillo2011,Shu2016b}. As a result, the inner mass budgets of close-pair deflectors, as well as the feasibility of measuring possible light--mass offsets in such systems, are still poorly constrained.

HSC~J0233$-$0205 was first identified as a grade-A strong-lens candidate \citep{Anton2020} in Hyper Suprime-Cam Subaru Strategic Program imaging (HSC-SSP; \citealt{Aihara2018}) by the Survey of Gravitationally Lensed Objects in HSC Imaging (SuGOHI; \citealt{Sonnenfeld2018}). HSC~J0233$-$0205 offers a rare opportunity to study the mass distribution of an extremely close galaxy-pair lens. The two deflecting galaxies in HSC~J0233$-$0205 are separated by only $\simeq3.6$ kpc in projection, placing the system among the most compact galaxy-pair lenses currently known. In addition, the lens redshift of $z_d\simeq0.8$ allows us to probe such a close projected pair at a relatively early cosmic epoch, when galaxy assembly, gas accretion, and merger-driven structural evolution were more active than in the local Universe.

In this work, we present the spectroscopic confirmation and initial lens modeling of HSC~J0233$-$0205 using ground-based imaging and spectroscopy. The main purpose of this study is to establish the system as a close galaxy-pair lens, obtain first-order constraints on its mass, light, and source properties, and highlight its potential as a valuable target for future high-resolution observations. Such follow-up data will be essential for more precise measurements of the inner mass distribution, dark-matter fraction, and possible light–mass offsets in this compact interacting lens system.

This paper is organized as follows. Section~\ref{sec:data} describes the HSC imaging, P200/DBSP spectroscopy, and redshift measurements. Section~\ref{sec:lens_modelling} presents the joint five-band lens modeling and comparison with previous results. Section~\ref{sec:diss} estimates the stellar masses and enclosed dark-matter fractions. Section~\ref{sec:con} summarizes the main results and discusses the implications of HSC~J0233$-$0205 as an extremely close galaxy-pair lens.
Throughout this Letter we adopt $\Omega_m=0.3$, $\Omega_\Lambda=0.7$, and $h=0.7$; at $z_d=0.79$, $1\arcsec$ corresponds to $7.476$ kpc. Magnitudes are in the AB system.

\section{Observations and Redshifts} \label{sec:data}

We use five-band HSC-SSP PDR3 imaging in the $g$, $r$, $i$, $z$, and $y$ bands \citep{Aihara2022}. We extract $39\times39$ pixel cutouts centered on the lens candidate at $\mathrm{RA}=38.34375^\circ$ and $\mathrm{Dec}=-2.09190^\circ$, corresponding to a field of view of $6\arcsec.55\times6\arcsec.55$ for the HSC pixel scale of $0\arcsec.168$. PSF models are generated at the target position using the HSC PDR3 PSF service.

\begin{figure*}
    \centering
    \includegraphics[scale=0.55]{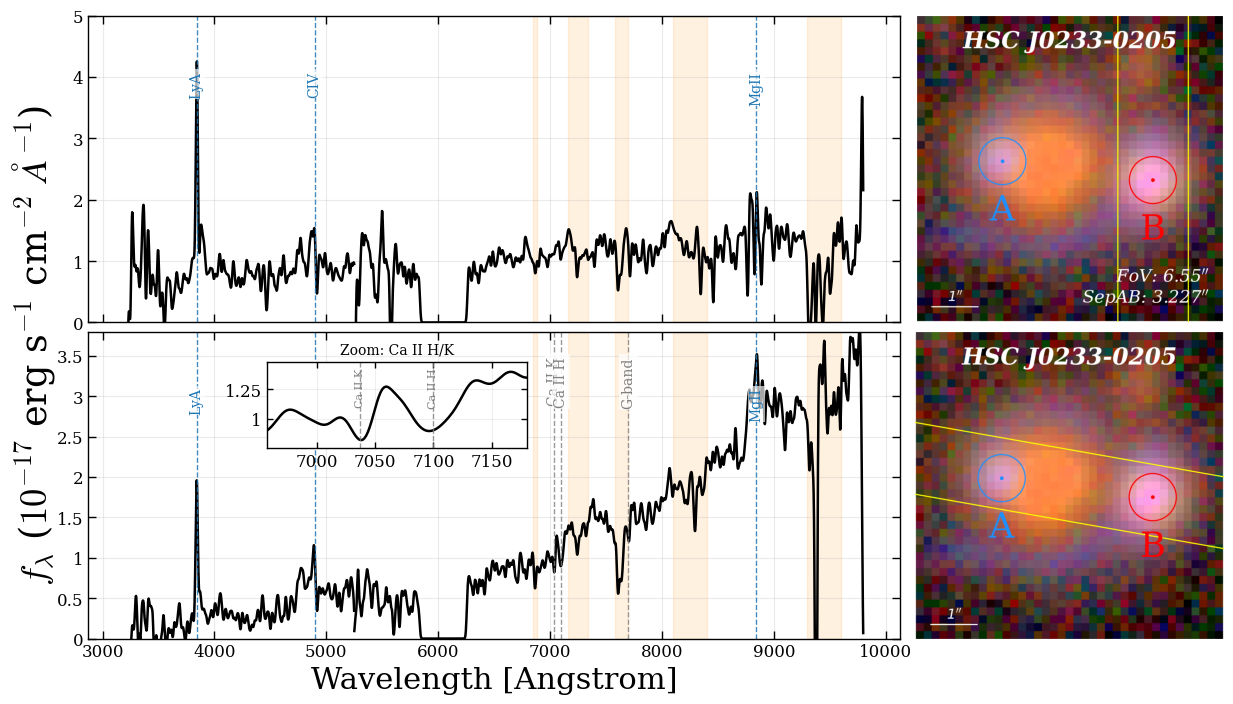}
    % \caption{P200/DBSP spectroscopy of HSC~J0233$-$0205. Top panels show the slit placements. The spectra of images A and B show the same source features at $z_s=2.160\pm0.002$, while the lens-galaxy absorption features give $z_d=0.790\pm0.022$.}
    \caption{
        P200/DBSP spectroscopy of HSC~J0233$-$0205. 
        The right panels show the HSC color image of the system, with the two multiple images labeled A and B. North is up and east is to the left.
        The lower-right panel shows the slit position used for the spectroscopic observation. 
        The left panels show the extracted one-dimensional spectra of image B (top) and the foreground lens galaxies plus image A (bottom). 
        The blue dashed lines mark emission lines from the lensed source at $z_s=2.160$, including Ly$\alpha$, C~IV, and Mg~II. 
        The gray dashed lines in the lower panel indicate absorption features associated with the lens galaxies at $z_d=0.790$, including Ca~II H/K and the G-band; the inset shows a zoom-in around the Ca~II H/K absorption doublet. The orange shaded bands mark major telluric absorption regions.
    }
    \label{fig:spec}
\end{figure*}

We observed HSC~J0233$-$0205 with P200/DBSP on the nights of 2023 October 15 and 16, obtaining one 1800 s exposure on each night with a $1\arcsec.5$ slit, both at an airmass of approximately 1.2. The first slit position covered image B, while the second covered images A and B and the lensing galaxies (Figure~\ref{fig:spec}). The spectra were reduced with our \textsc{Python} pipeline following standard steps, including bias subtraction, flat-fielding, cosmic-ray rejection, one-dimensional extraction, wavelength calibration, and flux calibration.

Images A and B show the same source emission features, including Ly$\alpha$, C\,\textsc{iv}, and Mg\,\textsc{ii}, yielding $z_s=2.160\pm0.002$ and confirming that they are multiple images of the same background source. Because image A is blended with the foreground lens galaxy, its spectrum contains contributions from both the lensed source and the deflector. To measure the deflector redshift, we therefore fit the lens-dominated portion of the image-A spectrum with the galaxy templates of \citet{Hutchinson2016}, masking source emission lines and telluric regions. The best-fitting redshift is $z_d=0.790\pm0.022$, with the model reproducing absorption features such as Ca\,\textsc{ii} H\&K and the G band. This spectroscopic redshift agrees well with the HSC DR3 photometric-redshift estimate, $z_{d,\rm phot}=0.790\pm0.048$ \citep[photoz\_best\_mizuki; see][for details]{Tanaka2018,Nishizawa2020}.

\section{Lens Modeling} \label{sec:lens_modelling}

We model the five HSC bands jointly with \texttt{lenstronomy} \citep{Birrer2018}. The mass distribution is represented by two elliptical power-law (EPL) components, one for each deflector, plus external shear. For each EPL component,
\begin{equation}
\Sigma(x,y)=\Sigma_{\rm crit}\,\frac{3-\gamma_{\rm EPL}}{2}
\left(\frac{\theta_{\rm E}}{\sqrt{q_{\rm EPL}x^2+y^2/q_{\rm EPL}}}\right)^{\gamma_{\rm EPL}-1} .
\end{equation}
The mass centroids are fixed to the corresponding lens-light centroids, as the current ground-based imaging data do not provide sufficient constraining power to robustly measure possible light--mass centroid offsets. The two deflectors are modeled with de~Vaucouleurs light profiles, with the S\'ersic index fixed to $n_s=4$. For each deflector, the centroid $(x,y)$, projected axis ratio $q_s$, and position angle $\phi_s$ are tied across bands, while the apparent magnitude $m$ and effective radius $r_{\rm e}$ are fitted independently in each band.

Motivated by the multi-component lensed morphology, we model the source-plane surface brightness phenomenologically with three S\'ersic components. We emphasize that this parameterization is designed to flexibly reproduce the source light distribution, and should not be interpreted as a unique decomposition into physically distinct galaxies. For each component, the centroid $(x,y)$, position angle $\phi_s$, and axis ratio $q_s$ are tied across all bands, while the magnitude $m$, effective radius $r_{\rm e}$, and S\'ersic index $n_s$ are fitted independently in each band, except for S\'ersic~3, for which $n_s$ is fixed to 4. S\'ersic~3 has $q_s=1$ and a very small $r_{\rm e}$, and therefore effectively represents a point-like source component. The magnification factor $\mu$ is computed from the best-fitting lens model in each band.

\begin{figure*}[ht!]
    \centering  
    \includegraphics[width=0.96\textwidth]{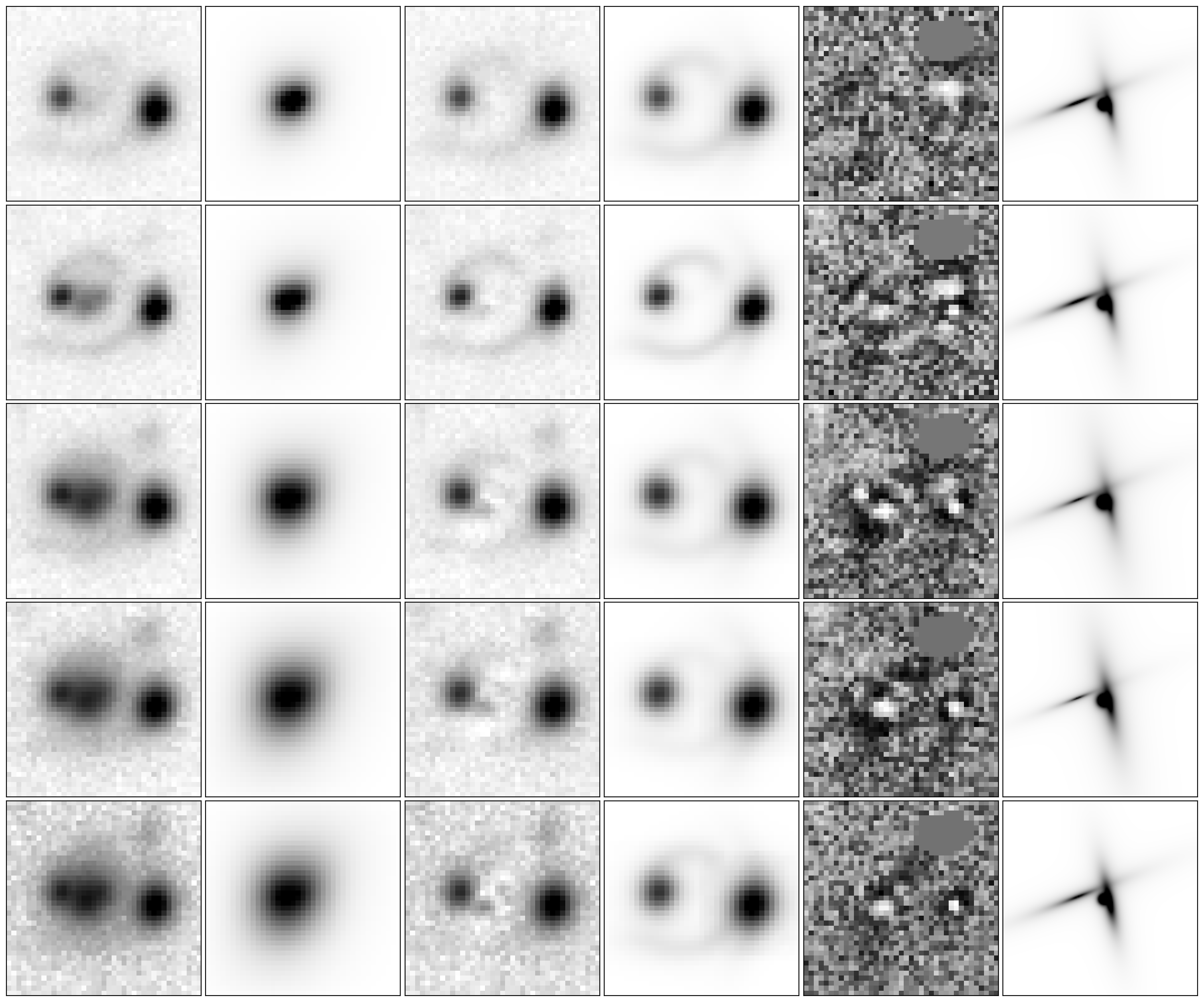}  
    \includegraphics[width=0.96\textwidth]{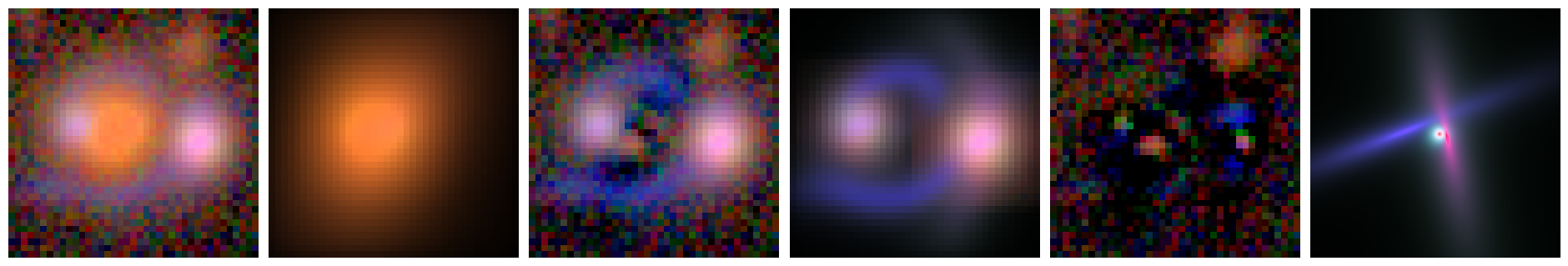}
    \caption{Best-fitting five-band HSC lens-modeling results. Columns show the data, lens-light model, lens-light-subtracted data, lensed-source model, residual, and source reconstruction. The display settings for the first to fifth columns are the same as those used in Figure~\ref{fig:spec}. The source-plane reconstructions in the sixth column cover a $3\arcsec \times 3\arcsec$ field of view with a pixel scale of $0\arcsec.0168$. The bottom row shows the color-composite model.}
    \label{fig:single_band_results}
\end{figure*}

\begin{table*}[]
\caption{Parameters of the EPL mass models.}
\label{tab:EPLresult}
\centering
\begin{tabular}{cccccc}
\hline\hline
$\theta_{\rm E}$ (\arcsec) & $\gamma_{\rm EPL}$        & $\phi_{\rm EPL}$ ($^\circ$) & $q_{\rm EPL}$             & $\gamma_{\rm shear,1}$                      & $\gamma_{\rm shear,2}$                     \\ \hline
$0.774\pm0.011$ & $2.051\pm0.018$ & $60.458\pm2.303$ & $0.559\pm0.030$ & \multirow{2}{*}{$0.034\pm0.005$} & \multirow{2}{*}{$0.122\pm0.007$} \\
$0.767\pm0.014$ & $1.746\pm0.012$ & $16.118\pm2.679$ & $0.679\pm0.020$ &                                             &                                            \\ \hline
\end{tabular}
\\
\textbf{Notes:} The lens mass centroids are fixed to the corresponding S\'ersic light centroids and are therefore not listed here. Position angles are defined with $\phi=0^\circ$ pointing west and increasing counterclockwise, so that north corresponds to $\phi=90^\circ$.
\end{table*}

\begin{table*}[]
\caption{Parameters of the lens light.}
\label{tab:lens_light}
\centering
\begin{tabular}{cccccccc}
\hline\hline
band & $m$ & center x (\arcsec) & center y (\arcsec) & $\phi_s (^\circ)$ & $q_s$ & $r_e$ (\arcsec) & $n_s$ \\
\hline

\multicolumn{8}{c}{\textbf{S\'ersic 1}} \\
$g$ & $23.983 \pm 0.048$ & \multirow{5}{*}{$-0.214 \pm 0.004$} & \multirow{5}{*}{$0.113 \pm 0.003$} & \multirow{5}{*}{$55.139 \pm 1.587$} & \multirow{5}{*}{$0.571 \pm 0.016$} & $0.408 \pm 0.031$ & $4.000$ \\
$r$ & $23.476 \pm 0.137$ &  & & & & $0.809 \pm 0.023$ & $4.000$ \\
$i$ & $22.064 \pm 0.179$ &  & & & & $0.744 \pm 0.024$ & $4.000$ \\
$z$ & $21.653 \pm 0.163$ &  & & & & $0.816 \pm 0.032$ & $4.000$ \\
$y$ & $21.205 \pm 0.157$ &  & & & & $0.723 \pm 0.016$ & $4.000$ \\
\hline

\multicolumn{8}{c}{\textbf{S\'ersic 2}} \\
$g$ & $24.401 \pm 0.097$ & \multirow{5}{*}{$-0.692 \pm 0.001$} & \multirow{5}{*}{$0.062 \pm 0.001$} & \multirow{5}{*}{$89.591 \pm 0.483$} & \multirow{5}{*}{$0.713 \pm 0.028$} & $0.781 \pm 0.014$ & $4.000$ \\
$r$ & $22.183 \pm 0.079$ &  & & & & $0.683 \pm 0.025$ & $4.000$ \\
$i$ & $20.481 \pm 0.082$ &  & & & & $0.658 \pm 0.023$ & $4.000$ \\
$z$ & $20.136 \pm 0.058$ &  & & & & $0.757 \pm 0.018$ & $4.000$ \\
$y$ & $19.615 \pm 0.065$ &  & & & & $0.665 \pm 0.018$ & $4.000$ \\
\hline
\end{tabular}
\textbf{Notes:} The two S\'ersic components describe the light distributions of the two foreground lens galaxies. Here $m$ is the apparent magnitude, $(x,y)$ is the centroid relative to the image-plane coordinate origin, $\phi_s$ is the position angle, $q_s$ is the projected axis ratio, $r_{\rm e}$ is the effective radius, and $n_s$ is the S\'ersic index. Position angles follow the same convention as in Table~\ref{tab:EPLresult}, and the coordinate origin is defined in the image plane (see Section~\ref{sec:data} for details).
\end{table*}

\begin{table*}[]
\caption{Parameters of the source light.}
\label{tab:source_light}
\centering
\begin{tabular}{ccccccccc}
\hline\hline
band & $m$ & $\mu$ & center x (\arcsec) & center y (\arcsec) & $\phi_s (^\circ)$ & $q_s$ & $r_e$ (\arcsec) & $n_s$ \\ 
\hline
\multicolumn{9}{c}{\textbf{S\'ersic 1}} \\
$g$  & $24.300 \pm 0.046$ & $17.498$ & \multirow{5}{*}{$-0.334 \pm 0.010$} & \multirow{5}{*}{$0.019 \pm 0.005$} & \multirow{5}{*}{$0.370 \pm 0.012$} & \multirow{5}{*}{$0.143 \pm 0.012$} & $0.616 \pm 0.018$ & $4.065 \pm 0.067$ \\
$r$  & $24.164 \pm 0.147$ & $17.484$ &  & & & & $0.359 \pm 0.041$ & $2.481 \pm 0.137$ \\
$i$  & $25.233 \pm 0.107$ & $17.059$ &  & & & & $0.600 \pm 0.036$ & $3.697 \pm 0.068$ \\
$z$  & $25.309 \pm 0.262$ & $17.655$ &  & & & & $0.508 \pm 0.025$ & $4.120 \pm 0.074$ \\
$y$  & $23.859 \pm 0.126$ & $17.376$ &  & & & & $0.475 \pm 0.022$ & $3.426 \pm 0.063$ \\
\hline

\multicolumn{9}{c}{\textbf{S\'ersic 2}} \\
$g$  & $24.356 \pm 0.065$ & $7.073$ & \multirow{5}{*}{$0.123 \pm 0.009$} & \multirow{5}{*}{$-0.050 \pm 0.005$} & \multirow{5}{*}{$178.679 \pm 0.017$} & \multirow{5}{*}{$0.218 \pm 0.013$} & $0.439 \pm 0.024$ & $4.691 \pm 0.092$ \\
$r$  & $23.843 \pm 0.137$ & $7.053$ &  & & & & $0.458 \pm 0.026$ & $5.244 \pm 0.084$ \\
$i$  & $23.421 \pm 0.133$ & $7.221$ &  & & & & $0.620 \pm 0.017$ & $5.214 \pm 0.066$ \\
$z$  & $23.317 \pm 0.156$ & $7.222$ &  & & & & $0.500 \pm 0.042$ & $5.056 \pm 0.099$ \\
$y$  & $19.887 \pm 0.264$ & $7.641$ &  & & & & $0.084 \pm 0.014$ & $5.833 \pm 0.110$ \\
\hline

\multicolumn{9}{c}{\textbf{S\'ersic 3}} \\
$g$  & $22.808 \pm 0.075$ & $3.431$ & \multirow{5}{*}{$0.041 \pm 0.007$} & \multirow{5}{*}{$-0.016 \pm 0.003$} & \multirow{5}{*}{$0.000 \pm 0.000$} & \multirow{5}{*}{$1.000 \pm 0.000$} & $0.007 \pm 0.001$ & $4.000$ \\
$r$  & $22.344 \pm 0.226$ & $3.290$ &  & & & & $0.008 \pm 0.001$ & $4.000$ \\
$i$  & $21.507 \pm 0.230$ & $3.388$ &  & & & & $0.007 \pm 0.001$ & $4.000$ \\
$z$  & $20.347 \pm 0.307$ & $3.472$ &  & & & & $0.006 \pm 0.001$ & $4.000$ \\
$y$  & $18.956 \pm 0.474$ & $3.627$ &  & & & & $0.004 \pm 0.001$ & $4.000$ \\
\hline
\end{tabular}
\textbf{Notes:} Same as Table~\ref{tab:lens_light}, but for the source-light model. $\mu$ denotes absolute magnification.
\end{table*}

The best-fitting model reproduces the five-band morphology well (Figure~\ref{fig:single_band_results}), with a reduced $\chi^2$ of 1.49. The posterior median parameters of the mass model, lens-light model, and source-light model are summarized in Tables~\ref{tab:EPLresult}, \ref{tab:lens_light}, and \ref{tab:source_light}, respectively. In comparison with the semi-automated GLEE-based \citep{Suyu2012} modeling of this system by \citet{Schuldt2023}, our normalized-residual maps show less apparent structure and no prominent ring-like residual features. Consistently, our fit yields a lower reduced $\chi^2$ than the value of 1.87 reported by \citet{Schuldt2023}. 

Based on the median mass model, the two lens galaxies have Einstein radii of $0\arcsec.774\pm0\arcsec.011$ and $0\arcsec.767\pm0\arcsec.014$ (Table~\ref{tab:EPLresult}), and their projected separation is $0\arcsec.481\pm0\arcsec.004$, or $3.596\pm0.046$ kpc. From the area enclosed by the critical curve, we obtain a circularized Einstein radius of $\theta_{\rm E}=1\arcsec.549\pm0\arcsec.018$ for the overall system. Integrating the convergence map within the critical curve gives $M_{\rm lensing,crit}=(9.626\pm0.010)\times10^{11}M_\odot$. Within apertures corresponding to the $z$-band effective radii, the enclosed lensing masses are $M_{\rm lensing,lens1}=(2.388\pm0.096)\times10^{11}M_\odot$ and $M_{\rm lensing,lens2}=(2.221\pm0.073)\times10^{11}M_\odot$. 

The reconstructed source is described by three S\'ersic components (Table~\ref{tab:source_light}). S\'ersic~1 and S\'ersic~2 are extended components with highly flattened morphologies, with axis ratios of $q_s=0.143\pm0.012$ and $0.218\pm0.013$, respectively. Their centroids are separated by $0\arcsec.462\pm0\arcsec.013$, corresponding to $3.830\pm0.108$ kpc at $z_s=2.160$, which may indicate that the background source consists of two closely separated star-forming components or galaxies. S\'ersic~3 is much more compact, with $r_{\rm e}\lesssim0\arcsec.01$ in all bands and $q_s$ fixed to unity, and is nearly aligned with S\'ersic~2. The inferred magnifications differ substantially among the three components, with S\'ersic~1 being the most strongly magnified ($\mu\simeq17$), S\'ersic~2 having intermediate magnification ($\mu\simeq7$), and the compact S\'ersic~3 component having lower magnification ($\mu\simeq3.3$--$3.6$).

\section{Stellar Masses and Dark-matter Fractions} \label{sec:diss}

\begin{figure*}
    \centering
    \includegraphics[scale=0.55]{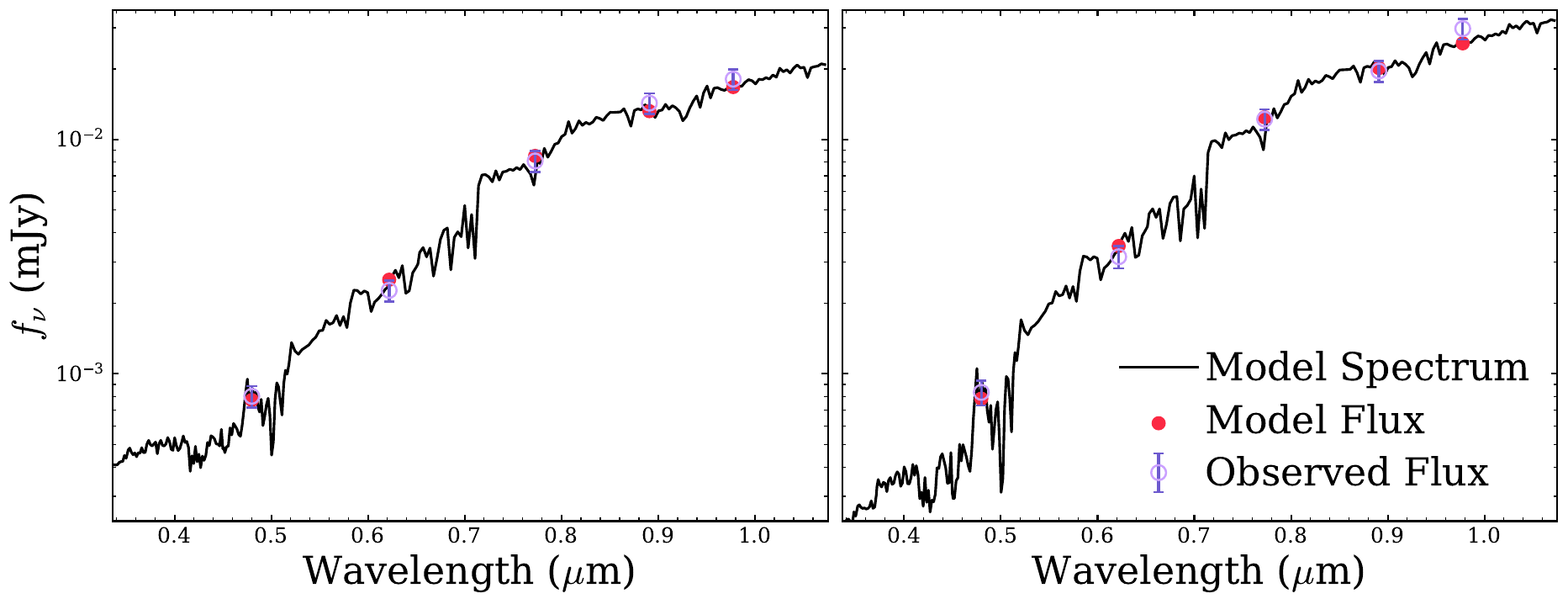}
    \caption{{\tt CIGALE} SED fitting results for the two lens galaxies. The black curves show the best-fitting model spectra, the red points show the model fluxes integrated through the HSC filters, and the open purple points show the observed lens photometry used in the fit.}
    \label{fig:sed}
\end{figure*}

We estimate the lens-galaxy stellar masses with \texttt{CIGALE} \citep{Boquien2019}, using the HSC five-band photometry, BC03 stellar-population models \citep{Bruzual2003}, and a Chabrier IMF \citep{Chabrier2003}. The inferred stellar masses are $M_{\ast,\mathrm{lens1}}=(0.854\pm0.212)\times10^{11}M_\odot$ and $M_{\ast,\mathrm{lens2}}=(1.524\pm0.296)\times10^{11}M_\odot$. The best-fitting SEDs from \texttt{CIGALE}, together with the observed HSC photometry, are shown in Figure~\ref{fig:sed}. The fitted effective radii suggest that both galaxies are relatively extended for early-type galaxies of similar stellar mass and redshift \citep{vanderWel2014}.

To estimate the projected dark-matter fraction, we use the same procedure for different apertures. 
For a given aperture, we first integrate the median lens-light model within that aperture and compare it with the total lens-light flux to obtain the enclosed light fraction, $f_\star(<r_{\rm ap})$. 
Assuming that stellar mass follows light, the enclosed stellar mass is then given by
$M_\ast(<r_{\rm ap})=f_\star(<r_{\rm ap})M_{\ast,\rm total}$. 
Combining this with the lensing mass measured from the convergence map in Section\,\ref{sec:lens_modelling}, $M_{\rm lensing}(<r_{\rm ap})$, we compute the dark-matter fraction as
\begin{equation}
 f_{\rm dm}(<r_{\rm ap}) = 1-\frac{M_\ast(<r_{\rm ap})}{M_{\rm lensing}(<r_{\rm ap})}.
\end{equation}

For the aperture enclosed by the critical curve, the enclosed light fraction is 
$f_{\star,\mathrm{crit}}=0.66$, yielding 
$M_{\ast,\mathrm{crit}}=(1.570\pm0.240)\times10^{11}M_\odot$. 
Together with the enclosed lensing mass 
$M_{\rm lensing,crit}=(9.626\pm0.010)\times10^{11}M_\odot$, this gives
$f_{\rm dm,crit}=83.7\pm2.5\%$. 
We further apply the same calculation within circular apertures defined by the $z$-band effective radius of each lens galaxy. By definition, each aperture encloses half of the corresponding S\'ersic component’s total light, so we adopt $f_{\star,r_e}=0.5$ for both lenses. The enclosed lensing masses are $M_{\rm lensing,lens1}=(2.388\pm0.096)\times10^{11}M_\odot$ and $M_{\rm lensing,lens2}=(2.221\pm0.073)\times10^{11}M_\odot$.
The resulting dark-matter fractions are 
$f_{\rm dm}(<r_e)=82.1\pm4.5\%$ for lens~1 and 
$f_{\rm dm}(<r_e)=65.7\pm6.7\%$ for lens~2. 

\begin{figure}
    \centering
    \includegraphics[scale=0.45]{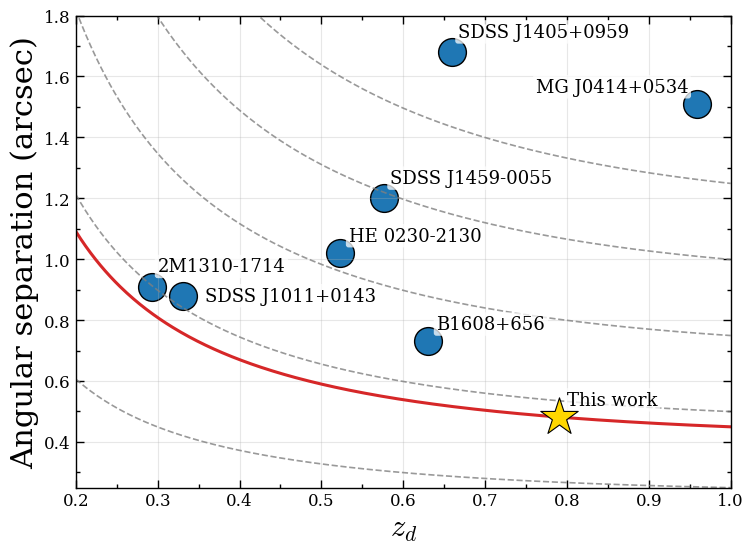}
    \caption{
Deflector redshift versus lens--lens angular separation for HSC~J0233$-$0205 and a literature sample of close-pair lens systems from \citet{Lucey2018,Shu2016b,Suyu2010a,Ertl2024,Willis2005,Rusu2014,Tonry1999}. The yellow star marks the system presented in this work. The gray dashed curves indicate constant projected physical separations of 2, 4, 6, 8, and 10 kpc, from bottom to top. The red curve shows the angular separation corresponding to the projected deflector separation of HSC~J0233$-$0205, $3.596$ kpc, as a function of deflector redshift.
}
    \label{fig:zd_vs_sep}
\end{figure}

For comparison with previous work, we compile known strong-lensing systems produced by close galaxy-pair deflectors, adopting a projected lens--lens separation of $<50$ kpc as the close-pair criterion \citep{Gonzalez2023}. We include only systems for which the lens--lens separation has been explicitly measured from observations. Among them, HSC~J0233$-$0205 has the smallest projected lens--lens separation and the second-highest deflector redshift. Figure~\ref{fig:zd_vs_sep} shows only systems with lens--lens angular separations smaller than $1\farcs8$; wider-separation systems, including CASSOWARY~5 \citep{Grillo2011}, SL2S~J08544$-$0121 \citep{Suyu2010b}, and DES~J0408$-$5354 \citep{Agnello2017}, are not plotted.

The present analysis is limited by the available ground-based imaging and by the lack of spatially resolved IFU spectroscopy. The current data do not allow us to robustly fit independent mass and light centroids, and the possible light--mass offset is therefore not measured in this work. High-resolution imaging would better reveal tidal features, improve the lens-light decomposition, and enable a direct search for offsets between luminous and dark-matter components. IFU observations would further provide spatially resolved stellar kinematics, help break degeneracies in the mass model, and potentially constrain the merger dynamics of the two deflectors, such as their relative line-of-sight velocity.

\section{Conclusions} \label{sec:con}

We have confirmed HSC~J0233$-$0205 as a strong-lensing system produced by a close galaxy pair at $z_d=0.790\pm0.022$ and lenses a multi-component background source at $z_s=2.160\pm0.002$. The deflectors are separated by only $3.596\pm0.046$ kpc, and the overall circularized Einstein radius is $1\arcsec.549\pm0\arcsec.018$. Joint five-band modeling gives two comparable Einstein radii, $0\arcsec.774\pm0\arcsec.011$ and $0\arcsec.767\pm0\arcsec.014$, and a total critical-curve mass of $(9.626\pm0.010)\times10^{11}M_\odot$. Combining lensing masses with SED-based stellar masses gives a projected dark-matter fraction of $83.7\pm2.5\%$ within the critical curve, and $82.1\pm4.5\%$ and $65.7\pm6.7\%$ within the $z$-band effective radii of the two deflectors. 

HSC~J0233$-$0205 is therefore a rare, extremely compact galaxy-pair lens and a promising target for future high-resolution imaging and spatially resolved spectroscopy. Future high-resolution imaging, including observations from upcoming surveys such as Euclid \citep{Euclid2025} and CSST \citep{CSST2026}, would enable more precise measurements of the dark-matter fraction, tests for possible offsets between the light and mass distributions, and searches for additional merger signatures such as tidal tails. If IFU observations become available, the internal kinematics of the two lens galaxies could also be studied in detail, providing a direct test of whether the close projected pair is a physically interacting system.

\section{Acknowledgments}
Z.H. acknowledges support from the National Natural Science Foundation of China (Grant No.~12403104). This research uses data obtained through the Telescope Access Program (TAP), which has been funded by the TAP association, including Centre for Astronomical Mega-Science CAS(CAMS), XMU, PKU, THU, USTC, NJU, YNU, and SYSU. We thank Houzun Chen for insightful discussions. 

The Hyper Suprime-Cam (HSC) collaboration includes the astronomical communities of Japan and Taiwan, and Princeton University. The HSC instrumentation and software were developed by the National Astronomical Observatory of Japan (NAOJ), the Kavli Institute for the Physics and Mathematics of the Universe (Kavli IPMU), the University of Tokyo, the High Energy Accelerator Research Organization (KEK), the Academia Sinica Institute for Astronomy and Astrophysics in Taiwan (ASIAA), and Princeton University. Funding was contributed by the FIRST program from Japanese Cabinet Office, the Ministry of Education, Culture, Sports, Science and Technology (MEXT), the Japan Society for the Promotion of Science (JSPS), Japan Science and Technology Agency (JST), the Toray Science Foundation, NAOJ, Kavli IPMU, KEK, ASIAA, and Princeton University.

Based [in part] on data collected at the Subaru Telescope and retrieved from the HSC data archive system, which is operated by Subaru Telescope and Astronomy Data Center at National Astronomical Observatory of Japan.

%% For this sample we use BibTeX plus aasjournalv7.bst to generate the

\bibliography{citations}{}

@ARTICLE{Lucey2018,
       author = {{Lucey}, John R. and {Schechter}, Paul L. and {Smith}, Russell J. and {Anguita}, T.},
        title = "{Serendipitous discovery of quadruply imaged quasars: two diamonds}",
      journal = {\mnras},
         year = 2018,
        month = may,
       volume = {476},
       number = {1},
        pages = {927-932},
          doi = {10.1093/mnras/sty243},
archivePrefix = {arXiv},
       eprint = {1711.02674},
 primaryClass = {astro-ph.GA},
       adsurl = {https://ui.adsabs.harvard.edu/abs/2018MNRAS.476..927L}
}

@ARTICLE{Ertl2024,
       author = {{Ertl}, S. and {Schuldt}, S. and {Suyu}, S.~H. and {Schechter}, P.~L. and {Halkola}, A. and {Wagner}, J.},
        title = "{The missing quasar image in the gravitationally lensed quasar HE0230{\ensuremath{-}}2130: Implications for the cored lens mass distribution and dark satellites}",
      journal = {\aap},
         year = 2024,
        month = may,
       volume = {685},
          eid = {A15},
        pages = {A15},
          doi = {10.1051/0004-6361/202347689},
archivePrefix = {arXiv},
       eprint = {2308.05181},
 primaryClass = {astro-ph.GA},
       adsurl = {https://ui.adsabs.harvard.edu/abs/2024A&A...685A..15E}
}

@ARTICLE{Rusu2014,
       author = {{Rusu}, Cristian E. and {Oguri}, Masamune and {Minowa}, Yosuke and {Iye}, Masanori and {More}, Anupreeta and {Inada}, Naohisa and {Oya}, Shin},
        title = "{Adaptive optics observations of the gravitationally lensed quasar SDSS J1405+0959}",
      journal = {\mnras},
         year = 2014,
        month = nov,
       volume = {444},
       number = {3},
        pages = {2561-2570},
          doi = {10.1093/mnras/stu1621},
archivePrefix = {arXiv},
       eprint = {1408.1425},
 primaryClass = {astro-ph.GA},
       adsurl = {https://ui.adsabs.harvard.edu/abs/2014MNRAS.444.2561R}
}

@ARTICLE{Agnello2017,
       author = {{Agnello}, A. and {Lin}, H. and {Buckley-Geer}, L. and {Treu}, T. and {Bonvin}, V. and {Courbin}, F. and {Lemon}, C. and {Morishita}, T. and {Amara}, A. and {Auger}, M.~W. and {Birrer}, S. and {Chan}, J. and {Collett}, T. and {More}, A. and {Fassnacht}, C.~D. and {Frieman}, J. and {Marshall}, P.~J. and {McMahon}, R.~G. and {Meylan}, G. and {Suyu}, S.~H. and {Castander}, F. and {Finley}, D. and {Howell}, A. and {Kochanek}, C. and {Makler}, M. and {Martini}, P. and {Morgan}, N. and {Nord}, B. and {Ostrovski}, F. and {Schechter}, P. and {Tucker}, D. and {Wechsler}, R. and {Abbott}, T.~M.~C. and {Abdalla}, F.~B. and {Allam}, S. and {Benoit-L{\'e}vy}, A. and {Bertin}, E. and {Brooks}, D. and {Burke}, D.~L. and {Rosell}, A. Carnero and {Kind}, M. Carrasco and {Carretero}, J. and {Crocce}, M. and {Cunha}, C.~E. and {D'Andrea}, C.~B. and {da Costa}, L.~N. and {Desai}, S. and {Dietrich}, J.~P. and {Eifler}, T.~F. and {Flaugher}, B. and {Fosalba}, P. and {Garc{\'\i}a-Bellido}, J. and {Gaztanaga}, E. and {Gill}, M.~S. and {Goldstein}, D.~A. and {Gruen}, D. and {Gruendl}, R.~A. and {Gschwend}, J. and {Gutierrez}, G. and {Honscheid}, K. and {James}, D.~J. and {Kuehn}, K. and {Kuropatkin}, N. and {Li}, T.~S. and {Lima}, M. and {Maia}, M.~A.~G. and {March}, M. and {Marshall}, J.~L. and {Melchior}, P. and {Menanteau}, F. and {Miquel}, R. and {Ogando}, R.~L.~C. and {Plazas}, A.~A. and {Romer}, A.~K. and {Sanchez}, E. and {Schindler}, R. and {Schubnell}, M. and {Sevilla-Noarbe}, I. and {Smith}, M. and {Smith}, R.~C. and {Sobreira}, F. and {Suchyta}, E. and {Swanson}, M.~E.~C. and {Tarle}, G. and {Thomas}, D. and {Walker}, A.~R.},
        title = "{Models of the strongly lensed quasar DES J0408-5354}",
      journal = {\mnras},
         year = 2017,
        month = dec,
       volume = {472},
       number = {4},
        pages = {4038-4050},
          doi = {10.1093/mnras/stx2242},
archivePrefix = {arXiv},
       eprint = {1702.00406},
 primaryClass = {astro-ph.GA},
       adsurl = {https://ui.adsabs.harvard.edu/abs/2017MNRAS.472.4038A}
}

@ARTICLE{Suyu2010a,
       author = {{Suyu}, S.~H. and {Marshall}, P.~J. and {Auger}, M.~W. and {Hilbert}, S. and {Blandford}, R.~D. and {Koopmans}, L.~V.~E. and {Fassnacht}, C.~D. and {Treu}, T.},
        title = "{Dissecting the Gravitational lens B1608+656. II. Precision Measurements of the Hubble Constant, Spatial Curvature, and the Dark Energy Equation of State}",
      journal = {\apj},
         year = 2010,
        month = mar,
       volume = {711},
       number = {1},
        pages = {201-221},
          doi = {10.1088/0004-637X/711/1/201},
archivePrefix = {arXiv},
       eprint = {0910.2773},
 primaryClass = {astro-ph.CO},
       adsurl = {https://ui.adsabs.harvard.edu/abs/2010ApJ...711..201S}
}

@ARTICLE{Spergel2000,
       author = {{Spergel}, David N. and {Steinhardt}, Paul J.},
        title = "{Observational Evidence for Self-Interacting Cold Dark Matter}",
      journal = {\prl},
         year = 2000,
        month = apr,
       volume = {84},
       number = {17},
        pages = {3760-3763},
          doi = {10.1103/PhysRevLett.84.3760}
}

@ARTICLE{Schuldt2023,
       author = {{Schuldt}, S. and {Suyu}, S.~H. and {Ca{\~n}ameras}, R. and {Shu}, Y. and {Taubenberger}, S. and {Ertl}, S. and {Halkola}, A.},
        title = "{HOLISMOKES. X. Comparison between neural network and semi-automated traditional modeling of strong lenses}",
      journal = {\aap},
         year = 2023,
        month = may,
       volume = {673},
          eid = {A33},
        pages = {A33},
          doi = {10.1051/0004-6361/202244534},
archivePrefix = {arXiv},
       eprint = {2207.10124},
 primaryClass = {astro-ph.CO},
       adsurl = {https://ui.adsabs.harvard.edu/abs/2023A&A...673A..33S}
}

@ARTICLE{vanderWel2014,
       author = {{vanderWel}, A. and {Franx}, M. and {van Dokkum}, P.~G. and {Skelton}, R.~E. and {Momcheva}, I.~G. and {Whitaker}, K.~E. and {Brammer}, G.~B. and {Bell}, E.~F. and {Rix}, H.-W. and {Wuyts}, S. and {Ferguson}, H.~C. and {Holden}, B.~P. and {Barro}, G. and {Koekemoer}, A.~M. and {Chang}, Yu-Yen and {McGrath}, E.~J. and {H{\"a}ussler}, B. and {Dekel}, A. and {Behroozi}, P. and {Fumagalli}, M. and {Leja}, J. and {Lundgren}, B.~F. and {Maseda}, M.~V. and {Nelson}, E.~J. and {Wake}, D.~A. and {Patel}, S.~G. and {Labb{\'e}}, I. and {Faber}, S.~M. and {Grogin}, N.~A. and {Kocevski}, D.~D.},
        title = "{3D-HST+CANDELS: The Evolution of the Galaxy Size-Mass Distribution since z = 3}",
      journal = {\apj},
         year = 2014,
        month = jun,
       volume = {788},
       number = {1},
          eid = {28},
        pages = {28},
          doi = {10.1088/0004-637X/788/1/28},
archivePrefix = {arXiv},
       eprint = {1404.2844},
 primaryClass = {astro-ph.GA},
       adsurl = {https://ui.adsabs.harvard.edu/abs/2014ApJ...788...28V}
}

@ARTICLE{Tanaka2018,
       author = {{Tanaka}, Masayuki and {Coupon}, Jean and {Hsieh}, Bau-Ching and {Mineo}, Sogo and {Nishizawa}, Atsushi J. and {Speagle}, Joshua and {Furusawa}, Hisanori and {Miyazaki}, Satoshi and {Murayama}, Hitoshi},
        title = "{Photometric redshifts for Hyper Suprime-Cam Subaru Strategic Program Data Release 1}",
      journal = {\pasj},
         year = 2018,
        month = jan,
       volume = {70},
          eid = {S9},
        pages = {S9},
          doi = {10.1093/pasj/psx077},
archivePrefix = {arXiv},
       eprint = {1704.05988},
 primaryClass = {astro-ph.GA},
       adsurl = {https://ui.adsabs.harvard.edu/abs/2018PASJ...70S...9T}
}

@ARTICLE{Nishizawa2020,
       author = {{Nishizawa}, Atsushi J. and {Hsieh}, Bau-Ching and {Tanaka}, Masayuki and {Takata}, Tadafumi},
        title = "{Photometric Redshifts for the Hyper Suprime-Cam Subaru Strategic Program Data Release 2}",
      journal = {arXiv e-prints},
         year = 2020,
        month = feb,
          eid = {arXiv:2003.01511},
        pages = {arXiv:2003.01511},
          doi = {10.48550/arXiv.2003.01511},
archivePrefix = {arXiv},
       eprint = {2003.01511},
 primaryClass = {astro-ph.GA},
       adsurl = {https://ui.adsabs.harvard.edu/abs/2020arXiv200301511N}
}

@ARTICLE{Auger2010,
       author = {{Auger}, M.~W. and {Treu}, T. and {Bolton}, A.~S. and {Gavazzi}, R. and {Koopmans}, L.~V.~E. and {Marshall}, P.~J. and {Moustakas}, L.~A. and {Burles}, S.},
        title = "{The Sloan Lens ACS Survey. X. Stellar, Dynamical, and Total Mass Correlations of Massive Early-type Galaxies}",
      journal = {\apj},
         year = 2010,
        month = nov,
       volume = {724},
       number = {1},
        pages = {511-525},
          doi = {10.1088/0004-637X/724/1/511},
archivePrefix = {arXiv},
       eprint = {1007.2880},
 primaryClass = {astro-ph.CO},
       adsurl = {https://ui.adsabs.harvard.edu/abs/2010ApJ...724..511A}
}

@ARTICLE{Chabrier2003,
       author = {{Chabrier}, Gilles},
        title = "{Galactic Stellar and Substellar Initial Mass Function}",
      journal = {\pasp},
         year = 2003,
        month = jul,
       volume = {115},
       number = {809},
        pages = {763-795},
          doi = {10.1086/376392},
archivePrefix = {arXiv},
       eprint = {astro-ph/0304382},
 primaryClass = {astro-ph},
       adsurl = {https://ui.adsabs.harvard.edu/abs/2003PASP..115..763C}
}

@ARTICLE{Bruzual2003,
       author = {{Bruzual}, G. and {Charlot}, S.},
        title = "{Stellar population synthesis at the resolution of 2003}",
      journal = {\mnras},
         year = 2003,
        month = oct,
       volume = {344},
       number = {4},
        pages = {1000-1028},
          doi = {10.1046/j.1365-8711.2003.06897.x},
archivePrefix = {arXiv},
       eprint = {astro-ph/0309134},
 primaryClass = {astro-ph},
       adsurl = {https://ui.adsabs.harvard.edu/abs/2003MNRAS.344.1000B}
}

@ARTICLE{Boquien2019,
       author = {{Boquien}, M. and {Burgarella}, D. and {Roehlly}, Y. and {Buat}, V. and {Ciesla}, L. and {Corre}, D. and {Inoue}, A.~K. and {Salas}, H.},
        title = "{CIGALE: a python Code Investigating GALaxy Emission}",
      journal = {\aap},
         year = 2019,
        month = feb,
       volume = {622},
          eid = {A103},
        pages = {A103},
          doi = {10.1051/0004-6361/201834156},
archivePrefix = {arXiv},
       eprint = {1811.03094},
 primaryClass = {astro-ph.GA},
       adsurl = {https://ui.adsabs.harvard.edu/abs/2019A&A...622A.103B}
}

@ARTICLE{Birrer2018,
       author = {{Birrer}, Simon and {Amara}, Adam},
        title = "{lenstronomy: Multi-purpose gravitational lens modelling software package}",
      journal = {Physics of the Dark Universe},
         year = 2018,
        month = dec,
       volume = {22},
        pages = {189-201},
          doi = {10.1016/j.dark.2018.11.002},
archivePrefix = {arXiv},
       eprint = {1803.09746},
 primaryClass = {astro-ph.CO},
       adsurl = {https://ui.adsabs.harvard.edu/abs/2018PDU....22..189B}
}

@ARTICLE{Shu2016b,
       author = {{Shu}, Yiping and {Bolton}, Adam S. and {Moustakas}, Leonidas A. and {Stern}, Daniel and {Dey}, Arjun and {Brownstein}, Joel R. and {Burles}, Scott and {Spinrad}, Hyron},
        title = "{Kiloparsec Mass/Light Offsets in the Galaxy Pair-Ly{\ensuremath{\alpha}} Emitter Lens System SDSS J1011+0143}",
      journal = {\apj},
         year = 2016,
        month = mar,
       volume = {820},
       number = {1},
          eid = {43},
        pages = {43},
          doi = {10.3847/0004-637X/820/1/43},
archivePrefix = {arXiv},
       eprint = {1602.02927},
 primaryClass = {astro-ph.GA},
       adsurl = {https://ui.adsabs.harvard.edu/abs/2016ApJ...820...43S}
}

@ARTICLE{Grillo2011,
       author = {{Grillo}, C. and {Christensen}, L.},
        title = "{Dark matter-rich early-type galaxies in the CASSOWARY 5 strong lensing system}",
      journal = {\mnras},
         year = 2011,
        month = dec,
       volume = {418},
       number = {2},
        pages = {929-937},
          doi = {10.1111/j.1365-2966.2011.19544.x},
archivePrefix = {arXiv},
       eprint = {1108.0678},
 primaryClass = {astro-ph.CO},
       adsurl = {https://ui.adsabs.harvard.edu/abs/2011MNRAS.418..929G}
}

@ARTICLE{Sonnenfeld2018,
       author = {{Sonnenfeld}, Alessandro and {Chan}, James H.~H. and {Shu}, Yiping and {More}, Anupreeta and {Oguri}, Masamune and {Suyu}, Sherry H. and {Wong}, Kenneth C. and {Lee}, Chien-Hsiu and {Coupon}, Jean and {Yonehara}, Atsunori and {Bolton}, Adam S. and {Jaelani}, Anton T. and {Tanaka}, Masayuki and {Miyazaki}, Satoshi and {Komiyama}, Yutaka},
        title = "{Survey of Gravitationally-lensed Objects in HSC Imaging (SuGOHI). I. Automatic search for galaxy-scale strong lenses}",
      journal = {\pasj},
         year = 2018,
        month = jan,
       volume = {70},
          eid = {S29},
        pages = {S29},
          doi = {10.1093/pasj/psx062},
archivePrefix = {arXiv},
       eprint = {1704.01585},
 primaryClass = {astro-ph.GA},
       adsurl = {https://ui.adsabs.harvard.edu/abs/2018PASJ...70S..29S}
}

@ARTICLE{Bolton2008,
       author = {{Bolton}, Adam S. and {Treu}, Tommaso and {Koopmans}, L{\'e}on V.~E. and {Gavazzi}, Rapha{\"e}l and {Moustakas}, Leonidas A. and {Burles}, Scott and {Schlegel}, David J. and {Wayth}, Randall},
        title = "{The Sloan Lens ACS Survey. VII. Elliptical Galaxy Scaling Laws from Direct Observational Mass Measurements}",
      journal = {\apj},
         year = 2008,
        month = sep,
       volume = {684},
       number = {1},
        pages = {248-259},
          doi = {10.1086/589989},
archivePrefix = {arXiv},
       eprint = {0805.1932},
 primaryClass = {astro-ph},
       adsurl = {https://ui.adsabs.harvard.edu/abs/2008ApJ...684..248B}
}

@ARTICLE{Aihara2018,
       author = {{Aihara}, Hiroaki and {Arimoto}, Nobuo and {Armstrong}, Robert and {Arnouts}, St{\'e}phane and {Bahcall}, Neta A. and {Bickerton}, Steven and {Bosch}, James and {Bundy}, Kevin and {Capak}, Peter L. and {Chan}, James H.~H. and {Chiba}, Masashi and {Coupon}, Jean and {Egami}, Eiichi and {Enoki}, Motohiro and {Finet}, Francois and {Fujimori}, Hiroki and {Fujimoto}, Seiji and {Furusawa}, Hisanori and {Furusawa}, Junko and {Goto}, Tomotsugu and {Goulding}, Andy and {Greco}, Johnny P. and {Greene}, Jenny E. and {Gunn}, James E. and {Hamana}, Takashi and {Harikane}, Yuichi and {Hashimoto}, Yasuhiro and {Hattori}, Takashi and {Hayashi}, Masao and {Hayashi}, Yusuke and {He{\l}miniak}, Krzysztof G. and {Higuchi}, Ryo and {Hikage}, Chiaki and {Ho}, Paul T.~P. and {Hsieh}, Bau-Ching and {Huang}, Kuiyun and {Huang}, Song and {Ikeda}, Hiroyuki and {Imanishi}, Masatoshi and {Inoue}, Akio K. and {Iwasawa}, Kazushi and {Iwata}, Ikuru and {Jaelani}, Anton T. and {Jian}, Hung-Yu and {Kamata}, Yukiko and {Karoji}, Hiroshi and {Kashikawa}, Nobunari and {Katayama}, Nobuhiko and {Kawanomoto}, Satoshi and {Kayo}, Issha and {Koda}, Jin and {Koike}, Michitaro and {Kojima}, Takashi and {Komiyama}, Yutaka and {Konno}, Akira and {Koshida}, Shintaro and {Koyama}, Yusei and {Kusakabe}, Haruka and {Leauthaud}, Alexie and {Lee}, Chien-Hsiu and {Lin}, Lihwai and {Lin}, Yen-Ting and {Lupton}, Robert H. and {Mandelbaum}, Rachel and {Matsuoka}, Yoshiki and {Medezinski}, Elinor and {Mineo}, Sogo and {Miyama}, Shoken and {Miyatake}, Hironao and {Miyazaki}, Satoshi and {Momose}, Rieko and {More}, Anupreeta and {More}, Surhud and {Moritani}, Yuki and {Moriya}, Takashi J. and {Morokuma}, Tomoki and {Mukae}, Shiro and {Murata}, Ryoma and {Murayama}, Hitoshi and {Nagao}, Tohru and {Nakata}, Fumiaki and {Niida}, Mana and {Niikura}, Hiroko and {Nishizawa}, Atsushi J. and {Obuchi}, Yoshiyuki and {Oguri}, Masamune and {Oishi}, Yukie and {Okabe}, Nobuhiro and {Okamoto}, Sakurako and {Okura}, Yuki and {Ono}, Yoshiaki and {Onodera}, Masato and {Onoue}, Masafusa and {Osato}, Ken and {Ouchi}, Masami and {Price}, Paul A. and {Pyo}, Tae-Soo and {Sako}, Masao and {Sawicki}, Marcin and {Shibuya}, Takatoshi and {Shimasaku}, Kazuhiro and {Shimono}, Atsushi and {Shirasaki}, Masato and {Silverman}, John D. and {Simet}, Melanie and {Speagle}, Joshua and {Spergel}, David N. and {Strauss}, Michael A. and {Sugahara}, Yuma and {Sugiyama}, Naoshi and {Suto}, Yasushi and {Suyu}, Sherry H. and {Suzuki}, Nao and {Tait}, Philip J. and {Takada}, Masahiro and {Takata}, Tadafumi and {Tamura}, Naoyuki and {Tanaka}, Manobu M. and {Tanaka}, Masaomi and {Tanaka}, Masayuki and {Tanaka}, Yoko and {Terai}, Tsuyoshi and {Terashima}, Yuichi and {Toba}, Yoshiki and {Tominaga}, Nozomu and {Toshikawa}, Jun and {Turner}, Edwin L. and {Uchida}, Tomohisa and {Uchiyama}, Hisakazu and {Umetsu}, Keiichi and {Uraguchi}, Fumihiro and {Urata}, Yuji and {Usuda}, Tomonori and {Utsumi}, Yousuke and {Wang}, Shiang-Yu and {Wang}, Wei-Hao and {Wong}, Kenneth C. and {Yabe}, Kiyoto and {Yamada}, Yoshihiko and {Yamanoi}, Hitomi and {Yasuda}, Naoki and {Yeh}, Sherry and {Yonehara}, Atsunori and {Yuma}, Suraphong},
        title = "{The Hyper Suprime-Cam SSP Survey: Overview and survey design}",
      journal = {PASJ},
         year = 2018,
        month = jan,
       volume = {70},
          eid = {S4},
        pages = {S4},
          doi = {10.1093/pasj/psx066},
archivePrefix = {arXiv},
       eprint = {1704.05858},
 primaryClass = {astro-ph.IM},
       adsurl = {https://ui.adsabs.harvard.edu/abs/2018PASJ...70S...4A}
}

@ARTICLE{Suyu2012,
       author = {{Suyu}, S.~H.},
        title = "{Cosmography from two-image lens systems: overcoming the lens profile slope degeneracy}",
      journal = {\mnras},
         year = 2012,
        month = oct,
       volume = {426},
       number = {2},
        pages = {868-879},
          doi = {10.1111/j.1365-2966.2012.21661.x},
archivePrefix = {arXiv},
       eprint = {1202.0287},
 primaryClass = {astro-ph.CO},
       adsurl = {https://ui.adsabs.harvard.edu/abs/2012MNRAS.426..868S}
}

@ARTICLE{Euclid2025,
       author = {{Euclid Collaboration} and {Mellier}, Y. and {Abdurro'uf} and {Acevedo Barroso}, J.~A. and {Ach{\'u}carro}, A. and {Adamek}, J. and {Adam}, R. and {Addison}, G.~E. and {Aghanim}, N. and {Aguena}, M. and {Ajani}, V. and {Akrami}, Y. and {Al-Bahlawan}, A. and {Alavi}, A. and {Albuquerque}, I.~S. and {Alestas}, G. and {Alguero}, G. and {Allaoui}, A. and {Allen}, S.~W. and {Allevato}, V. and {Alonso-Tetilla}, A.~V. and {Altieri}, B. and {Alvarez-Candal}, A. and {Alvi}, S. and {Amara}, A. and {Amendola}, L. and {Amiaux}, J. and {Andika}, I.~T. and {Andreon}, S. and {Andrews}, A. and {Angora}, G. and {Angulo}, R.~E. and {Annibali}, F. and {Anselmi}, A. and {Anselmi}, S. and {Arcari}, S. and {Archidiacono}, M. and {Aric{\`o}}, G. and {Arnaud}, M. and {Arnouts}, S. and {Asgari}, M. and {Asorey}, J. and {Atayde}, L. and {Atek}, H. and {Atrio-Barandela}, F. and {Aubert}, M. and {Aubourg}, E. and {Auphan}, T. and {Auricchio}, N. and {Aussel}, B. and {Aussel}, H. and {Avelino}, P.~P. and {Avgoustidis}, A. and {Avila}, S. and {Awan}, S. and {Azzollini}, R. and {Baccigalupi}, C. and {Bachelet}, E. and {Bacon}, D. and {Baes}, M. and {Bagley}, M.~B. and {Bahr-Kalus}, B. and {Balaguera-Antolinez}, A. and {Balbinot}, E. and {Balcells}, M. and {Baldi}, M. and {Baldry}, I. and {Balestra}, A. and {Ballardini}, M. and {Ballester}, O. and {Balogh}, M. and {Ba{\~n}ados}, E. and {Barbier}, R. and {Bardelli}, S. and {Baron}, M. and {Barreiro}, T. and {Barrena}, R. and {Barriere}, J.-C. and {Barros}, B.~J. and {Barthelemy}, A. and {Bartolo}, N. and {Basset}, A. and {Battaglia}, P. and {Battisti}, A.~J. and {Baugh}, C.~M. and {Baumont}, L. and {Bazzanini}, L. and {Beaulieu}, J.-P. and {Beckmann}, V. and {Belikov}, A.~N. and {Bel}, J. and {Bellagamba}, F. and {Bella}, M. and {Bellini}, E. and {Benabed}, K. and {Bender}, R. and {Benevento}, G. and {Bennett}, C.~L. and {Benson}, K. and {Bergamini}, P. and {Bermejo-Climent}, J.~R. and {Bernardeau}, F. and {Bertacca}, D. and {Berthe}, M. and {Berthier}, J. and {Bethermin}, M. and {Beutler}, F. and {Bevillon}, C. and {Bhargava}, S. and {Bhatawdekar}, R. and {Bianchi}, D. and {Bisigello}, L. and {Biviano}, A. and {Blake}, R.~P. and {Blanchard}, A. and {Blazek}, J. and {Blot}, L. and {Bosco}, A. and {Bodendorf}, C. and {Boenke}, T. and {B{\"o}hringer}, H. and {Boldrini}, P. and {Bolzonella}, M. and {Bonchi}, A. and {Bonici}, M. and {Bonino}, D. and {Bonino}, L. and {Bonvin}, C. and {Bon}, W. and {Booth}, J.~T. and {Borgani}, S. and {Borlaff}, A.~S. and {Borsato}, E. and {Bose}, B. and {Botticella}, M.~T. and {Boucaud}, A. and {Bouche}, F. and {Boucher}, J.~S. and {Boutigny}, D. and {Bouvard}, T. and {Bouwens}, R. and {Bouy}, H. and {Bowler}, R.~A.~A. and {Bozza}, V. and {Bozzo}, E. and {Branchini}, E. and {Brando}, G. and {Brau-Nogue}, S. and {Brekke}, P. and {Bremer}, M.~N. and {Brescia}, M. and {Breton}, M.-A. and {Brinchmann}, J. and {Brinckmann}, T. and {Brockley-Blatt}, C. and {Brodwin}, M. and {Brouard}, L. and {Brown}, M.~L. and {Bruton}, S. and {Bucko}, J. and {Buddelmeijer}, H. and {Buenadicha}, G. and {Buitrago}, F. and {Burger}, P. and {Burigana}, C. and {Busillo}, V. and {Busonero}, D. and {Cabanac}, R. and {Cabayol-Garcia}, L. and {Cagliari}, M.~S. and {Caillat}, A. and {Caillat}, L. and {Calabrese}, M. and {Calabro}, A. and {Calderone}, G. and {Calura}, F. and {Camacho Quevedo}, B. and {Camera}, S. and {Campos}, L. and {Ca{\~n}as-Herrera}, G. and {Candini}, G.~P. and {Cantiello}, M. and {Capobianco}, V. and {Cappellaro}, E. and {Cappelluti}, N. and {Cappi}, A. and {Caputi}, K.~I. and {Cara}, C. and {Carbone}, C. and {Cardone}, V.~F. and {Carella}, E. and {Carlberg}, R.~G. and {Carle}, M. and {Carminati}, L. and {Caro}, F. and {Carrasco}, J.~M. and {Carretero}, J. and {Carrilho}, P. and {Carron Duque}, J. and {Carry}, B.},
        title = "{Euclid: I. Overview of the Euclid mission}",
      journal = {\aap},
         year = 2025,
        month = may,
       volume = {697},
          eid = {A1},
        pages = {A1},
          doi = {10.1051/0004-6361/202450810},
archivePrefix = {arXiv},
       eprint = {2405.13491},
 primaryClass = {astro-ph.CO},
       adsurl = {https://ui.adsabs.harvard.edu/abs/2025A&A...697A...1E}
}

@ARTICLE{CSST2026,
       author = {{CSST Collaboration} and {Gong}, Yan and {Miao}, Haitao and {Zhan}, Hu and {Li}, Zhao-Yu and {Shangguan}, Jinyi and {Li}, Haining and {Liu}, Chao and {Chen}, Xuefei and {Yuan}, Haibo and {Zhou}, Jilin and {Liu}, Hui-Gen and {Yu}, Cong and {Ji}, Jianghui and {Qi}, Zhaoxiang and {Liu}, Jiacheng and {Dai}, Zigao and {Wang}, Xiaofeng and {Zheng}, Zhenya and {Hao}, Lei and {Dou}, Jiangpei and {Ao}, Yiping and {Lin}, Zhenhui and {Zhang}, Kun and {Wang}, Wei and {Sun}, Guotong and {Li}, Ran and {Li}, Guoliang and {Xu}, Youhua and {Li}, Xinfeng and {Li}, Shengyang and {Wu}, Peng and {Zhang}, Jiuxing and {Wang}, Bo and {Bai}, Jinming and {Cai}, Yi-Fu and {Cai}, Zheng and {Cao}, Jie and {Chan}, Kwan Chuen and {Chang}, Jin and {Chen}, Xiaodian and {Chen}, Xuelei and {Chen}, Yuqin and {Chen}, Yun and {Cui}, Wei and {Dong}, Subo and {Du}, Pu and {Duan}, Wenying and {Fan}, Junhui and {Fan}, LuLu and {Fan}, Zhou and {Fan}, Zuhui and {Fang}, Taotao and {Fu}, Jianning and {Fu}, Liping and {Fu}, Zhensen and {Gao}, Jian and {Gu}, Shenghong and {Gu}, Yidong and {Guo}, Qi and {Han}, Zhanwen and {Hu}, Bin and {Huang}, Zhiqi and {Ho}, Luis C. and {Jiang}, Linhua and {Jiang}, Ning and {Jing}, Yipeng and {Kang}, Xi and {Kong}, Xu and {Li}, Cheng and {Li}, Chengyuan and {Li}, Di and {Li}, Jing and {Li}, Nan and {Li}, Yang A. and {Liao}, Shilong and {Lin}, Weipeng and {Liu}, Fengshan and {Liu}, Jifeng and {Liu}, Xiangkun and {Liu}, Zhuokai and {Mao}, Ruiqing and {Mao}, Shude and {Meng}, Xianmin and {Pang}, Xiaoying and {Peng}, Xiyan and {Peng}, Yingjie and {Shan}, Huanyuan and {Shen}, Juntai and {Shen}, Shiyin and {Shen}, Zhiqiang and {Shi}, Sheng-Cai and {Shi}, Yong and {Tan}, Siyuan and {Tian}, Hao and {Wang}, Jianmin and {Wang}, Jun-Xian and {Wang}, Xin and {Wang}, Yuting and {Wu}, Hong and {Wu}, Jingwen and {Wu}, Xuebing and {Xu}, Chun and {Xue}, Xiang-Xiang and {Xue}, Yongquan and {Yang}, Ji and {Yang}, Xiaohu and {Yao}, Qijun and {Yuan}, Fangting and {Yuan}, Zhen and {Zhang}, Jun and {Zhang}, Pengjie and {Zhang}, Tianmeng and {Zhang}, Wei and {Zhang}, Xin and {Zhao}, Gang and {Zhao}, Gongbo and {Zhong}, Hongen and {Zhong}, Jing and {Zhou}, Liyong and {Zhu}, Wei and {Zu}, Ying},
        title = "{Introduction to the Chinese Space Station Survey Telescope (CSST)}",
      journal = {Science China Physics, Mechanics, and Astronomy},
         year = 2026,
        month = jan,
       volume = {69},
       number = {3},
          eid = {239501},
        pages = {239501},
          doi = {10.1007/s11433-025-2809-0},
archivePrefix = {arXiv},
       eprint = {2507.04618},
 primaryClass = {astro-ph.IM},
       adsurl = {https://ui.adsabs.harvard.edu/abs/2026SCPMA..6939501C}
}

@ARTICLE{Gonzalez2023,
       author = {{Gonzalez}, E.~J. and {Rodriguez}, F. and {Navarro-Giron{\'e}s}, D. and {Gazta{\~n}aga}, E. and {Siudek}, M. and {Garc{\'\i}a Lambas}, D. and {O'Mill}, A.~L. and {Renard}, P. and {Cabayol}, L. and {Carretero}, J. and {Casas}, R. and {De Vicente}, J. and {Eriksen}, M. and {Fernandez}, E. and {Garcia-Bellido}, J. and {Hildebrandt}, H. and {Miquel}, R. and {Padilla}, C. and {Sanchez}, E. and {Sevilla-Noarbe}, I. and {Tallada-Cresp{\'\i}}, P. and {Wittje}, A.},
        title = "{The PAU survey: close galaxy pairs identification and analysis}",
      journal = {\mnras},
         year = 2023,
        month = jul,
       volume = {522},
       number = {4},
        pages = {5655-5668},
          doi = {10.1093/mnras/stad1350},
archivePrefix = {arXiv},
       eprint = {2305.01952},
 primaryClass = {astro-ph.GA},
       adsurl = {https://ui.adsabs.harvard.edu/abs/2023MNRAS.522.5655G}
}

@ARTICLE{Tonry1999,
       author = {{Tonry}, John L. and {Kochanek}, Christopher S.},
        title = "{Redshifts of the Gravitational Lenses MG 0414+0534 and MG 0751+2716}",
      journal = {\aj},
         year = 1999,
        month = may,
       volume = {117},
       number = {5},
        pages = {2034-2038},
          doi = {10.1086/300834},
archivePrefix = {arXiv},
       eprint = {astro-ph/9809063},
 primaryClass = {astro-ph},
       adsurl = {https://ui.adsabs.harvard.edu/abs/1999AJ....117.2034T}
}

@ARTICLE{Suyu2010b,
       author = {{Suyu}, S.~H. and {Halkola}, A.},
        title = "{The halos of satellite galaxies: the companion of the massive elliptical lens SL2S J08544-0121}",
      journal = {\aap},
         year = 2010,
        month = dec,
       volume = {524},
          eid = {A94},
        pages = {A94},
          doi = {10.1051/0004-6361/201015481},
archivePrefix = {arXiv},
       eprint = {1007.4815},
 primaryClass = {astro-ph.CO},
       adsurl = {https://ui.adsabs.harvard.edu/abs/2010A&A...524A..94S}
}

@ARTICLE{Willis2005,
       author = {{Willis}, J.~P. and {Hewett}, P.~C. and {Warren}, S.~J.},
        title = "{The discovery of two new galaxy-galaxy lenses from the Sloan Digital Sky Survey}",
      journal = {\mnras},
         year = 2005,
        month = nov,
       volume = {363},
       number = {4},
        pages = {1369-1375},
          doi = {10.1111/j.1365-2966.2005.09533.x},
archivePrefix = {arXiv},
       eprint = {astro-ph/0508430},
 primaryClass = {astro-ph},
       adsurl = {https://ui.adsabs.harvard.edu/abs/2005MNRAS.363.1369W}
}

@ARTICLE{Koopmans2006,
       author = {{Koopmans}, L{\'e}on V.~E. and {Treu}, Tommaso and {Bolton}, Adam S. and {Burles}, Scott and {Moustakas}, Leonidas A.},
        title = "{The Sloan Lens ACS Survey. III. The Structure and Formation of Early-Type Galaxies and Their Evolution since z \raisebox{-0.5ex}\textasciitilde 1}",
      journal = {\apj},
         year = 2006,
        month = oct,
       volume = {649},
       number = {2},
        pages = {599-615},
          doi = {10.1086/505696},
archivePrefix = {arXiv},
       eprint = {astro-ph/0601628},
 primaryClass = {astro-ph},
       adsurl = {https://ui.adsabs.harvard.edu/abs/2006ApJ...649..599K}
}

@ARTICLE{Treu2010,
       author = {{Treu}, Tommaso},
        title = "{Strong Lensing by Galaxies}",
      journal = {\araa},
         year = 2010,
        month = sep,
       volume = {48},
        pages = {87-125},
          doi = {10.1146/annurev-astro-081309-130924},
archivePrefix = {arXiv},
       eprint = {1003.5567},
 primaryClass = {astro-ph.CO},
       adsurl = {https://ui.adsabs.harvard.edu/abs/2010ARA&A..48...87T}
}

@ARTICLE{Keeton2003,
       author = {{Keeton}, Charles R. and {Winn}, Joshua N.},
        title = "{The Quintuple Quasar: Mass Models and Interpretation}",
      journal = {\apj},
         year = 2003,
        month = jun,
       volume = {590},
       number = {1},
        pages = {39-51},
          doi = {10.1086/374833},
archivePrefix = {arXiv},
       eprint = {astro-ph/0302366},
 primaryClass = {astro-ph},
       adsurl = {https://ui.adsabs.harvard.edu/abs/2003ApJ...590...39K}
}

@article{Anton2020,
    author = {Jaelani, Anton T and More, Anupreeta and Oguri, Masamune and Sonnenfeld, Alessandro and Suyu, Sherry H and Rusu, Cristian E and Wong, Kenneth C and Chan, James H H and Kayo, Issha and Lee, Chien-Hsiu and Chao, Dani C-Y and Coupon, Jean and Inoue, Kaiki T and Futamase, Toshifumi},
    title = {Survey of Gravitationally lensed Objects in HSC Imaging (SuGOHI) – V. Group-to-cluster scale lens search from the HSC–SSP Survey},
    journal = {Monthly Notices of the Royal Astronomical Society},
    volume = {495},
    number = {1},
    pages = {1291-1310},
    year = {2020},
    month = {04},
    issn = {0035-8711},
    doi = {10.1093/mnras/staa1062},
    url = {https://doi.org/10.1093/mnras/staa1062},
    eprint = {https://academic.oup.com/mnras/article-pdf/495/1/1291/33290106/staa1062.pdf},}

@ARTICLE{Aihara2022,
       author = {{Aihara}, Hiroaki and {AlSayyad}, Yusra and {Ando}, Makoto and {Armstrong}, Robert and {Bosch}, James and {Egami}, Eiichi and {Furusawa}, Hisanori and {Furusawa}, Junko and {Harasawa}, Sumiko and {Harikane}, Yuichi and {Hsieh}, Bau-Ching and {Ikeda}, Hiroyuki and {Ito}, Kei and {Iwata}, Ikuru and {Kodama}, Tadayuki and {Koike}, Michitaro and {Kokubo}, Mitsuru and {Komiyama}, Yutaka and {Li}, Xiangchong and {Liang}, Yongming and {Lin}, Yen-Ting and {Lupton}, Robert H. and {Lust}, Nate B. and {MacArthur}, Lauren A. and {Mawatari}, Ken and {Mineo}, Sogo and {Miyatake}, Hironao and {Miyazaki}, Satoshi and {More}, Surhud and {Morishima}, Takahiro and {Murayama}, Hitoshi and {Nakajima}, Kimihiko and {Nakata}, Fumiaki and {Nishizawa}, Atsushi J. and {Oguri}, Masamune and {Okabe}, Nobuhiro and {Okura}, Yuki and {Ono}, Yoshiaki and {Osato}, Ken and {Ouchi}, Masami and {Pan}, Yen-Chen and {Plazas Malag{\'o}n}, Andr{\'e}s A. and {Price}, Paul A. and {Reed}, Sophie L. and {Rykoff}, Eli S. and {Shibuya}, Takatoshi and {Simunovic}, Mirko and {Strauss}, Michael A. and {Sugimori}, Kanako and {Suto}, Yasushi and {Suzuki}, Nao and {Takada}, Masahiro and {Takagi}, Yuhei and {Takata}, Tadafumi and {Takita}, Satoshi and {Tanaka}, Masayuki and {Tang}, Shenli and {Taranu}, Dan S. and {Terai}, Tsuyoshi and {Toba}, Yoshiki and {Turner}, Edwin L. and {Uchiyama}, Hisakazu and {Vijarnwannaluk}, Bovornpratch and {Waters}, Christopher Z. and {Yamada}, Yoshihiko and {Yamamoto}, Naoaki and {Yamashita}, Takuji},
        title = "{Third data release of the Hyper Suprime-Cam Subaru Strategic Program}",
      journal = {\pasj},
         year = 2022,
        month = apr,
       volume = {74},
       number = {2},
        pages = {247-272},
          doi = {10.1093/pasj/psab122},
archivePrefix = {arXiv},
       eprint = {2108.13045},
 primaryClass = {astro-ph.IM},
       adsurl = {https://ui.adsabs.harvard.edu/abs/2022PASJ...74..247A}
}

@ARTICLE{Granata2023,
       author = {{Granata}, G. and {Bergamini}, P. and {Grillo}, C. and {Meneghetti}, M. and {Mercurio}, A. and {Me{\v{s}}tri{\'c}}, U. and {Ragagnin}, A. and {Rosati}, P. and {Caminha}, G.~B. and {Tortorelli}, L. and {Vanzella}, E.},
        title = "{Exploring the low-mass regime of galaxy-scale strong lensing: Insights into the mass structure of cluster galaxies}",
      journal = {\aap},
         year = 2023,
        month = nov,
       volume = {679},
          eid = {A124},
        pages = {A124},
          doi = {10.1051/0004-6361/202347521},
archivePrefix = {arXiv},
       eprint = {2310.02310},
 primaryClass = {astro-ph.GA},
       adsurl = {https://ui.adsabs.harvard.edu/abs/2023A&A...679A.124G}
}

@ARTICLE{Gavazzi2007,
       author = {{Gavazzi}, Rapha{\"e}l and {Treu}, Tommaso and {Rhodes}, Jason D. and {Koopmans}, L{\'e}on V.~E. and {Bolton}, Adam S. and {Burles}, Scott and {Massey}, Richard J. and {Moustakas}, Leonidas A.},
        title = "{The Sloan Lens ACS Survey. IV. The Mass Density Profile of Early-Type Galaxies out to 100 Effective Radii}",
      journal = {\apj},
         year = 2007,
        month = sep,
       volume = {667},
        pages = {176-190},
          doi = {10.1086/519237},
archivePrefix = {arXiv},
       eprint = {astro-ph/0701589},
       adsurl = {https://ui.adsabs.harvard.edu/abs/2007ApJ...667..176G}
}
\bibliographystyle{aasjournalv7}



\end{document}